\documentclass{ws-rv961x669}
\usepackage{footmisc}
\usepackage{subfigure}   % required only when side-by-side / subfigures are used
\usepackage{ws-rv-thm}   % comment this line when `amsthm / theorem / ntheorem` package is used
\usepackage{ws-rv-van}   % numbered citation & references (default)
\usepackage{ws-index}   % to produce multiple indexes
\makeindex
\usepackage{graphicx}
\usepackage{subfigure}
\usepackage{textcomp}
\usepackage{slashed}

\def\ring#1{{\mathaccent'27 #1}}

\def\cri{{\ring{c}}}

\def\Eri{{\ring{E}}}

\def\ccfc#1#2{\cri^{(#1)}_{#2}}

\def\beq{\begin{equation}}
\def\eeq{\end{equation}}
\def\theta{\vartheta}

\newcommand{\ba}{\begin{eqnarray}}
\newcommand{\ea}{\end{eqnarray}}
\newcommand{\lsim}   {\mathrel{\mathop{\kern 0pt \rlap
  {\raise.2ex\hbox{$<$}}}
  \lower.9ex\hbox{\kern-.190em $\sim$}}}
\newcommand{\gsim}   {\mathrel{\mathop{\kern 0pt \rlap
  {\raise.2ex\hbox{$>$}}}
  \lower.9ex\hbox{\kern-.190em $\sim$}}}

\begin{document}
%\begin{chapter}
%\chapter[Using World Scientific's Review Volume Document Style]{Using World Scientific's Review Volume\\ Document Style\label{ra_ch1}}

%\author[F. Author and S. Author]{First Author and Second Author\footnote{Author footnote.}}
%\index[aindx]{Author, F.} % or \aindx{Author, F.}
%\index[aindx]{Author, S.} % or \aindx{Author, S.}

%\address{World Scientific Publishing Co, Production Department,\\
%5 Toh Tuck Link, Singapore 596224, \\
%f\_author@wspc.com.sg\footnote{Affiliation footnote.}}
%\markboth{Testing LIV with Neutrinos}{Floyd W. Stecker}

\title{Neutrino Physics and Astrophysics}

\chapter[Testing LIV with Neutrinos]{Testing Lorentz Invariance with Neutrinos}

\author[Floyd W. Stecker]{Floyd W. Stecker\footnote{Floyd.W.Stecker@nasa.gov}}
%\aindx{Author, F.}                       % author index entry
\address{NASA Goddard Space Flight Center \\ Greenbelt, MD 20771 USA \\
and University of California Los Angeles \\ Los Angeles, CA 90095 }
%Floyd.W.Stecker@nasa.gov}

%\end{chapter}
%\begin{abstract}
%The abstract should ...
%\end{abstract}
\vspace{1cm}
%\begin{appendix}[Optional Appendix Title]
%\section{Sample Appendix}
%Text...
%\end{appendix}
{\it Today we say that the law of relativity is supposed to be true at all energies, but someday somebody may come along and say how stupid we were. We do not know where we are stupid until we stick our neck out...And the only way to find out that we are wrong is to find out what our predictions are. It is absolutely necessary to make constructs.}

- Richard Feynman

\vspace{1cm}

\begin{abstract}

\noindent The search for a  theory that unifies general relativity and quantum theory has focused attention on models of physics at the Planck scale. One possible consequence of models
such as string theory may be that Lorentz invariance is not an exact symmetry of nature. 
We discuss here some possible experimental and observational tests of Lorentz invariance
involving neutrino physics and astrophysics.

\vspace{1cm}
 
\end{abstract}

\newpage

\tableofcontents

\newpage

\section{Introduction}\label{sec1.1}

Modern theoretical physics stands on two strong legs. General relativity, which sprang from the mind of Albert Einstein over a century ago, has been crucial to a description of physics on the largest scales. Quantum field theory, developed in the middle of the 20th century, has led to the modern view of physics on extremely small scales, allowing a detailed description of the interactions of the subatomic particles. It has also played a crucial role in the macroscopic world, giving physicists a framework for the description of emergent phenomena in condensed matter. However, a unification of these two fundamental branches of modern physics has been heretofore illusive. 

Even more problematic, the two theories appear to be incompatible, particularly failing at the nexus point, known as the Planck scale of $\lambda_{Pl} = \sqrt{G\hbar/c^3} \sim 10^{-35}$ m.~\cite{Planck:1899pl}. 
It is at this scale that the Compton wavelength and the Schwarzschild radius of a particle of mass $m_{Planck}$ are equal. This implies that for intervals at or near the Planck scale, the distortion of space-time geometry owing to quantum effects becomes catastrophic. Thus, we say that at this scale space-time physics as we know it "breaks down"~\cite{Mead:1964zz}. More likely, it is a red flag of our ignorance. We may say that the big-bang started at the Planck scale, but we really don't know what happened at that time. There has even been speculation of a pre-Plankian or trans-Planckian era~\cite{Stecker:1979ke,Gasperini:1992em} that influenced our present cosmology~\cite{Danielsson:2002kx}.

So what about physics at the Planck scale? In their efforts to provide a UV (i.e., high energy) completion for a quantum theory of general relativity, many theorists postulate drastic modifications to space-time at the Planck scale (e.g., Ref.~\refcite{Doplicher:1994tu}). Two examples of this are the introduction of "extra dimensions" and the postulation of a fundamental discreteness of space-time, with the building blocks of nature being extended objects. At stake is nothing less than our understanding of the nature of space and time on which all of physics is fashioned.
 
One possible modification to space-time structure that has received quite a bit of attention is the idea that the Lorentz space-time symmetry of relativity is not an exact symmetry of nature. Such a proposal is rather conservative as compared with other quantum gravity ideas.  Historically the symmetry groups that have been used to model physical phenomena have inevitably evolved over time, with Galilean symmetry being replaced by Lorentz symmetry and other particle physics symmetries being broken. Lorentz symmetry violation has been explored within string theory, loop quantum gravity, Ho\v{r}ava-Lifshitz gravity, causal dynamical triangulations, non-commutative geometry, doubly special relativity, among others (See, e.g., Refs.~\refcite{Mattingly:2005re} and~\refcite{Liberati:2013xla} and references therein).

An example of the importance of neutrino physics is the long recognized importance of the role of neutrino masses and oscillations in cosmology, in particular, their role in determining the large scale structure of the Universe~\cite{Stecker:1982dr,Shafi:1984ek}. Neutrino studies and observations can also provide sensitive tests of Lorentz invariance violation, which may be a the result of quantum gravity physics (QG). Studies of neutrino oscillations as well as observations of astrophysical neutrinos can be important in this regard.

\subsection{Neutrinos and Tests of Lorentz Invariance Violation}

Empirical studies of physical phenomena involving neutrinos, both in the laboratory and from cosmic sources, can be a useful probe in searching for new physics. In this chapter we will discuss the prospect of using observations of high energy neutrinos  produced in astrophysical sources in order to search for traces of physics at the Planck-scale. 
Such a project is advantageous for several reasons. Cosmic high energy neutrinos are unaffected by magnetic fields and effectively unaffected by interactions with matter. They can therefore reach us from cosmologically great distances, thus providing an extremely long baseline for probing the smallest deviations from Lorentz invariance and standard model physics. Because of the astronomically long baselines and very high energies of such neutrinos, they are therefore ideal probes for testing Lorentz and $\cal{CPT}$ symmetry as well as other new physics. In addition, considerations based on dimensional analysis indicate that LIV effects should generally increase with energy.

In the following sections, we will examine some of the observational consequences of LIV in the neutrino sector within the context of effective field theory (EFT), one that holds up to some limiting high energy scale that we take to be the Planck scale. 

\section{Effective Field Theories and Neutrino Physics}

While it is not possible to directly investigate space-time physics at the Planck energy of $\sim 10^{19}$ \ GeV, many lower energy testable effects have been predicted to arise from the violation of Lorentz invariance (LIV) at or near the Planck scale. The subject of investigating LIV has therefore generated much interest in the particle physics and astrophysics communities. Here we choose to discuss LIV within the useful framework of effective field theories (EFT).\footnote{There are non-EFT scenarios for either violating or modifying Lorentz invariance (see, e.g., Refs.~\refcite{Mattingly:2005re},~\refcite{Liberati:2013xla},~\refcite{Amelino-Camelia:2008aez} and references therein). However, these models do not easily lend themselves to particle physics tests as the dynamics of particle interactions is less well understood. We will not consider such scenarios here.} 
An effective field theory is a theory that is incomplete, but one that is an excellent approximation below a certain limiting energy scale. An example of an EFT is the Fermi four-fermion theory of weak interactions~\cite{Fermi:1933jpa} that holds up to an energy $\sim$ 100 GeV. This theory was replaced by the exact electroweak theory of Glashow, Weinberg and Salam. (See, e.g., Ref.~\refcite{Ryder:1985wq}). Using EFT methods allows one to
pose questions involving LIV that can have well-defined empirical answers without having the knowledge of an exact theory.

In the localized quantum field theory formalism the evolution of the state of a quantum system is determined by a unitary Hermitian operator that is a functional of a Lagrangian density, colloquially called {\it the Lagrangian}~\cite{Schwinger:1951xk,Schwinger:1953tb}.
The effect of incorporating possible new physics in such an EFT framework is accomplished by employing new operators and incorporating new free parameters into the Lagrangian of the theory with such operators that being constructed by identifying the relevant fields and symmetries that determine the possible new physics. 
Using such an effective field theory formalism, Lorentz invariance violation can be incorporated by the addition of new terms in the free particle Lagrangian that explicitly break Lorentz invariance. Since it is well known that Lorentz invariance holds quite well at accelerator energies, the extra Lorentz violating terms in the effective field theory Lagrangian must be very small up to such energies. 

We will assume that the EFT that breaks the symmetry of Lorentz invariance is a good approximation to a "true theory" and is one that holds up to the order of the Planck scale. Within the context of such an EFT, one can postulate the existence of additional terms in free particle Lagrangians that break Lorentz invariance, some of which violate $\cal{CPT}$ invariance. Within this EFT framework one can then describe effects arising from Planck-scale physics that can be manifested at lower energies,  however being suppressed at lower energies by terms in the Lagrangians inversely proportional to powers of the Planck mass. Such an EFT formalism was proposed with in the context of string theory~\cite{Kostelecky:1988zi,Kostelecky:1991ak}. In the rest of this chapter we will discuss the empirical effects of an EFT that incorporates operators in the Lagrangian that violate both Lorentz invariance and $\cal{CPT}$ invariance in the neutrino sector. In particular, we will place limits on a class of non-renormalizable, mass-dimension five and six Lorentz invariance violating operators that may be the result of Planck-scale physics. (See Appendix A for a discussion of Lagrangians and mass dimensions in EFT.)

 \section{Free particle propagation and modified kinematics}
\label{form}

LIV modifications to the Lagrangian can affect neutrino physics in various ways. We will consider here only two of these consequences, viz., (1) their effect on modifying neutrino oscillations and (2) the effect of resulting changes in the kinematics of particle interactions. Such modifications give rise to an energy dependent effective neutrino mass, and so change the patterns of neutrino oscillations.  They also introduce corrections to the matrix elements for existing interactions as well as create new interactions between standard model fermions.  However, for our purposes, the most important effect these terms have is to change the kinematics of particle interactions, leaving unchanged the matrix elements governed by existing standard model physics. 

Such changes can modify the threshold energies for particle interactions, allowing or forbidding such interactions~\cite{Coleman:1998ti,Stecker:2001vb,Altschul:2021wrm}. Since the Lorentz violating operators change the free field behavior and dispersion relation, interactions such as fermion-antifermion pair emission by neutrinos become kinematically allowed~\cite{Coleman:1998ti,Stecker:2001vb} and can cause significant observational effects if the neutrinos are slightly superluminal. Absent a violation of Lorentz invariance, such interactions are forbidden by conservation of energy and momentum.  

We now set up a simplified formalism to calculate the observational effect of these two specific anomalous interactions on the neutrino spectrum seen in IceCube.

\subsection{Mass dimension [d] = 4 LIV with rotational symmetry}

For an introduction demonstrating how LIV terms affect particle kinematics,
we consider the simple example of a free scalar particle Lagrangian with an additional small dimension-4 Lorentz violating term, assuming rotational symmetry~\cite{Coleman:1998ti}.  
\begin{equation}
\Delta \mathcal{L}_f = \partial_{i}\Phi^{*}{\bf \epsilon}\partial^{i}\Phi.
\end{equation}
which is not Lorentz invariant because it only involves spacial derivatives.

This modified Lagrangian leads to a propagator for a particle of mass $m$
\begin{equation}
-iD^{-1}~=~ (p_{(4)}^2~-~m^2)~+~\epsilon p^2.
\end{equation}
so that we obtain the dispersion relation
\begin{equation}
p_{(4)}^2~=~E^2~-~p^2~ \Rightarrow~ m^2 ~ +~ \epsilon p^2.
\label{LIVdispersion}
\end{equation}

In this example, the low energy "speed of light" maximum attainable particle velocity, here equal to 1 by convention, is replaced by a new maximum attainable velocity (MAV) as $v_{MAV} \neq 1$, which is 
changed by 
\begin{equation}
\delta v \equiv \delta = \epsilon/2.
\label{deltadef}
\end{equation}

Then,
\begin{equation}
{{\partial  E}\over{\partial |\vec{p}|}}  = {{|\vec{p}|}  \over {\sqrt
{|\vec{p}|^2 + m^2 v_{MAV} ^2}}} v_{MAV},
\label{groupvel}
\end{equation}
\vspace{12pt}
\noindent which goes  to $v_{MAV}$ at relativistic energies, $|\vec{p}|^2 \gg m^2$.

For the [d] = 4 case, the superluminal velocity of particle $I$ that is produced by the existence of one or more LIV terms in the free particle Lagrangian will be denoted by
\begin{equation}
v_{I, MAV} \equiv 1 + \delta_{I}
\label{v}
\end{equation}
We will always be in the relativistic limit $|\vec{p}|^2 \gg m^2$ for both neutrinos and electrons. Thus, the neutrino or electron velocity is just given by equation~(\ref{v})~\cite{Coleman:1998ti,Altschul:2021wrm}. 

\subsection{Fermion LIV operators with $[d] > 4$ LIV with rotational symmetry in the "Standard Model Extension" EFT}
\label{SME} 

In considering Lagrangians with $[d] > 4$, we can employ the EFT framework known as the "Standard Model Extension" (SME)~\cite{Kostelecky:1988zi}. This formalism was inspired by string theory and invokes small terms in the Lagrangian beyond the standard model that violate Lorentz invariance and $\cal{CPT}$ invariance while keeping the internal gauge symmetries of the standard model in the Lagrangian. (See appendix B.) As in the [d] = 4 case above~\cite{Coleman:1998ti}, we will consider here only the isotropic terms in the SME formalism and relate our treatment to those SME parameters in the conclusion. This "spherical cow", plain vanilla approximation is justified by our present lack of empirical
knowledge.\footnote{The only additional LIV terms in the free particle Lagrangians that we consider here are assumed to be rotationally invariant. The "standard model extension" (SME) EFT includes hundreds of possible anisotropic terms as well~\cite{Kostelecky:1988zi,Colladay:1998fq,Kostelecky:2009zp,Kostelecky:2011gq}. Since there is presently no observational evidence for any LIV, we chose the simplest isotropic approach here.  We assume that the LIV terms are isotropic in the rest system of the 2.73 K cosmic background radiation (CBR), a preferred system picked out by the universe itself. In actuality, we are moving with respect to this system by a velocity $\sim 10^{-3}$ of the speed of light (hereafter taking $c = \hbar = 1$).} 

In the cases where $[d] > 4$ Planck-suppressed operators dominate, there will be LIV terms that are proportional to $(E/M_{Pl})^n$, where $n = [d] - 4$, leading to values of $\delta_{I}$ that are energy dependent and are taken to be suppressed by appropriate powers of the Planck mass (see appendix B). 

If we again assume rotational invariance in the rest frame of the CBR and consider only the effects of Lorentz violation on freely propagating cosmic neutrinos. Thus, we only need to examine Lorentz violating modifications to the neutrino kinetic terms. Majorana neutrino couplings are ruled out in SME in the case of rotational symmetry~\cite{Kostelecky:2011gq}. Therefore we only consider Dirac neutrinos. 

A useful simplified formalism for analyzing Lorentz-modified kinematics, one that highlights the physical processes, is to wrap the additional Lorentz violating terms into an effective mass term, $\tilde{m}_I(E)$, which is the right hand side of equation (\ref{LIVdispersion}) labeled by a particle species index $I$.  We can further identify $\tilde{m}(E)$ using equation (\ref{LIVdispersion}), yielding the relation

\begin{equation} 
\tilde{m}_{I}^2(E)=m_{I}^2+ 2\delta_I E^2,
\label{effectivemass}
\end{equation}
where the velocity parameters $\delta_I$ are now energy dependent dimensionless  coefficients for each species that can be directly identified from the fundamental parameters in the Lagrangian. We have implicitly assumed that  $\delta_I$ is defined relative the velocity of light, $c = 1$. We extend this  definition by defining the parameter $\delta_{IJ} \equiv \delta_{I} - \delta_{J}$ as the Lorentz violating difference between the velocities of particles $I$ and $J$. In general $\delta_{IJ}$ will therefore be of the form
\begin{equation}
\delta_{IJ}=\sum_{n=0,1,2}\kappa_{IJ,n} \left(\frac{E}{M_{Pl}}\right)^n.
\label{d}
\end{equation}

Assuming the dominance of Planck-mass suppressed terms in the Lagrangian as tracers of Planck scale physics, it follows that that $\kappa_{\nu e,0} = 0$. We will also assume that 
$\kappa_{\nu e,0} \ll \kappa_{\nu e,1}, \kappa_{\nu e,2}$. Constraints are therefore most directly expressed in terms of limits on $\kappa_{\nu e,1}$ and $\kappa_{\nu e,2}$. For $[d] > 4$ the superluminal velocity excesses are given as integral multiples of $\kappa_{\nu e,1}$ and $\kappa_{\nu e,2}$ through the group velocity relation given by equation~(\ref{groupvel}). The $\kappa_{\nu e,0}$ coefficient has been tightly constrained from observations of extraterrestrial PeV neutrinos by the IceCube collaboration~\cite{Stecker:2013jfa,Stecker:2014xja}.\footnote{Several mechanisms have been proposed for the suppression of the LIV $[d] = 4$ term in the Lagrangian. See, e.g., the review in Ref.~\refcite{Liberati:2013xla}.}
\footnote{In equations (\ref{d}) and (\ref{sub}) we have not designated a helicity index on the $\kappa$ coefficients. The fundamental parameters in the Lagrangian are generally helicity dependent~\cite{Jacobson:2002ye}. In the $n = 1$ case a helicity dependence must be generated in the electron sector due to the $\cal{CPT}$ odd nature of the LIV term. However, the constraints on the electron coefficient are extremely tight from observations of the Crab nebula (see above, see also Ref.~\refcite{Jacobson:2003bn}). Thus, the contribution to $\kappa_{\nu e,1}$ from the electron sector can be neglected. In the $n = 2$ case, which is $\cal{CPT}$ even, we can set the left and right handed electron coefficients to be equal by imposing parity symmetry~\cite{Stecker:2014oxa}.}

There is an important connection between LIV and $\cal{CPT}$ violation. Whereas a local interacting theory that violates $\cal{CPT}$ invariance will also violate Lorentz invariance, the converse does not follow; an interacting theory that violates Lorentz invariance may, or may not, violate $\cal{CPT}$ invariance~\cite{Greenberg:2002uu,Greenberg:2003nv}. $\cal{CPT}$ is a symmetry that is conserved in a field theory formulation that satisfies the following three assumptions: (1) locality, (2) Lorentz invariance, and (3) hermiticity of the Hamiltonian. As an example, if strings, not points, are the fundamental elements of physics, $\cal{CPT}$ invariance can be violated because the locality condition of the $\cal{CPT}$ theorem~\cite{Schwinger:1951xk} does not hold~\cite{Kostelecky:1991ak,Lehnert:2016zym}.

In the framework of the standard model extension (SME) formalism~\cite{Colladay:1998fq} LIV terms of even mass dimension $[d]= 4 + n$, are $\cal{CPT}$-even and do not violate $\cal{CPT}$, whereas LIV terms of odd mass dimension are $\cal{CPT}$-odd and violate $\cal{CPT}$~\cite{Kostelecky:2009zp}. We can then specify a dominant term for $\delta_{IJ}$ in equation (\ref{d}) depending on our choice of $\cal{CPT}$. Thus, in considering Planck-mass suppressed LIV terms, the dominant term that admits $\cal{CPT}$ violation is the $n = 1$ term in equation (\ref{d}). On the other hand, if we require $\cal{CPT}$ conservation, the $n = 2$ term in equation (\ref{d}) is the dominant term. Therefore, we can choose as a good approximation to equation (\ref{d}), a single dominant term with one particular power of $n$ by specifying whether we are considering $\cal{CPT}$ even or odd LIV. We can then reduce the sum of the terms in the expansion given by equation (\ref{d}) to the leading terms only. As a result, $\delta_{IJ}$ reduces to
\begin{equation}
\delta_{IJ}~ \equiv \kappa_{IJ,n} \left({{E}\over{{M_{Pl}}}}\right)^{n} 
\label{sub}
\end{equation}
\vspace{12pt}
with $n = 1$ or $n = 2$ depending on the status of $\cal{CPT}$.

It will also be important later on to note that in the SME formalism, since odd-[d] LIV operators are $\cal{CPT}$ odd, the $\cal{CPT}$-conjugation property implies that neutrinos can be superluminal while antineutrinos are subluminal or vice versa~\cite{Kostelecky:2011gq}. This will have consequences in interpreting the results given in the later sections.
~
\begin{figure}%[tb]
 \begin{center}
  \includegraphics[width=8cm]{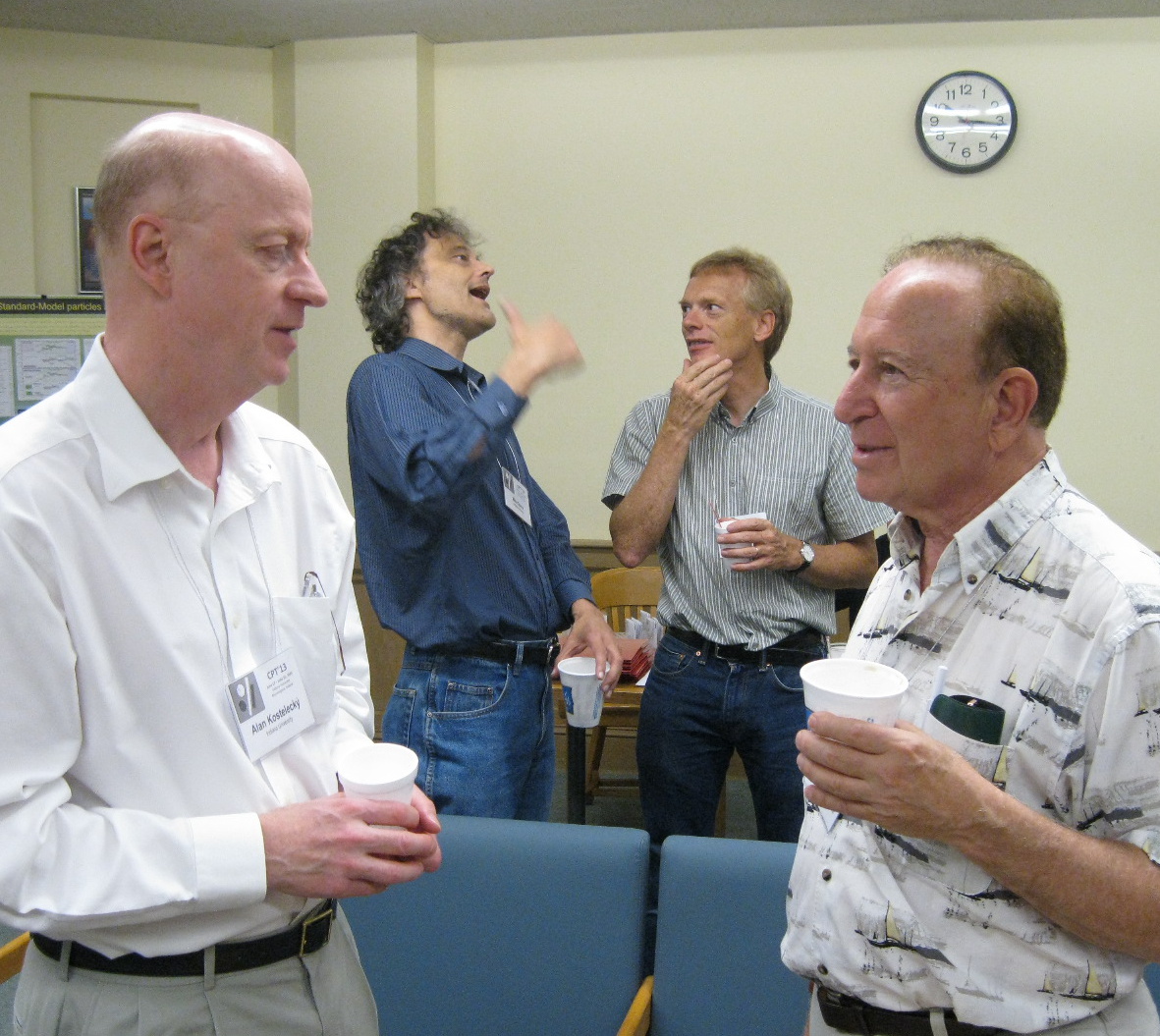}
  \caption{Alan Kosteleck\'y and the author (left to right in the foreground) at the CPT 2013 Symposium.
  In the background are Mike Snow (waving) and Rob Potting (Courtesy of Neil Russell) }\label{CPT13}
\end{center}
\end{figure}
 
\section{LIV in the neutrino sector I - Neutrino Oscillations}
\label{osc}

Let us now consider the effect on neutrino oscillations of Lorentz violating terms in the Lagrangian that are suppressed by powers of the Planck mass. We again note that the dominant term that admits $\cal{CPT}$ violation is the $n = 1$ term in equation (\ref{d}); the dominant term that conserves $\cal{CPT}$ is the $n = 2$ term in equation (\ref{d}). Given Planck mass suppression, we choose one of these two terms to be the single dominant term with one particular power of $n$, depending on whether $\cal{CPT}$ is conserved or not.  As a result, $\delta_{IJ}$ is given by equation (\ref{sub}). We take the effective mass, $\tilde{m}$, to be given by equation (\ref{effectivemass}) where the $m_i$ denotes one of the three possible mass eigenstates of the neutrino.

We then consider a neutrino with flavor $I$ with  momentum $p$ transitioning into flavor $J$. The amplitude for this neutrino to be in a mass eigenstate $i$ is then denoted by the matrix $U_{Ii}$ where ${\bf U}$ denotes the unitary matrix, $\sum U^{\dag}_{Ji}U_{Ii} = \partial_{IJ}$, where here the symbol $\partial_{IJ}$ denotes the usual mathematical delta function to distinguish it from to the velocity difference $\delta_{IJ}$ as defined in equation (\ref{d}). 

These considerations change the usual relations for neutrino oscillations. For example, in the case of atmospheric $\nu_{\mu}$ oscillations, the survival probability in the Lorentz invariant case is determined by the time evolution of the mass eigenstates and is given by 

\begin{equation}
P_{\nu_{\mu}} \ \simeq \ 1 - sin^2(2\theta_{23})sin^2\left({\Delta m^2_{atm}L}\over{4E}\right)
\label{atmLI}
\end{equation}
where $\Delta m^2 \ \equiv m_{i}^{2}$ -  $m_{j}^{2}$, so that the oscillation periods are determined by parameters involving the differences between
the squares of the neutrino mass eigenstates. Thus, the individual neutrino masses
themselves are not determined by the oscillations.

We can include the effects of LIV terms in the Lagrangian by making the substitutions
\begin{equation}
m^2_{atm} \ \rightarrow \ \ \tilde{m}^2_{atm}(E) \ =  \ m^2_{atm} + 2\delta_{ij} E^2
\end{equation}
\newline
\noindent where $\tilde{m}^2(E)$ is given by equation (\ref{effectivemass}). In that case, equation (\ref{atmLI}) is modified by an additional LIV term proportional to the difference in neutrino velocities, $\delta_{IJ}$. It immediately follows that
\begin{equation}
P_{\nu_{\mu}} \ \simeq \ 1 - sin^2(2\theta_{23})sin^2\left({{\Delta m^2_{atm}L}\over{4E}} + 
{{\delta_{IJ}}}{{EL\over{2}}}\right)
\label{atmLIV}
\end{equation}
\newline
\noindent where the square of the difference between the mass eigenvalues $\Delta m^2_{atm} = \Delta m^2_{31}$~\cite{Coleman:1998ti,Diaz:2011ia,Maccione:2011fr}. Recent results on atmospheric neutrino oscillations \cite{Gonzalez-Garcia:2004pka,Super-Kamiokande:2014exs,deSalas:2020pgw} (also Gonzalez-Garcia and Maltoni, private communication) give an upper limit for the difference in velocities between the $\nu_{\mu}$ and $\nu_{\tau}$ neutrinos, $\delta_{\nu_{\mu} \nu_{\tau}} < \cal{O}$ $(10^{-26})$. 

It is interesting to note that if Lorentz invariance is violated, equation (\ref{atmLIV}) implies that neutrino oscillations would occur even if neutrinos are massless or if the square of their mass differences is zero.   

We note that the LIV term in equation (\ref{atmLIV}) is linear in energy. Therefore, it can dominate at very high energies and very large distances~\cite{}, as is the case for many astrophysical applications. This dominance can be even more profound for high mass dimensions where $\delta_{IJ}$ is given by equation (\ref{sub}).

\section{Velocity difference between neutrinos and photons from TXS 0506+056}

The detection of a neutrino event detected by the IceCube collaboration
in temporal and spacial coincidence with a $\gamma$-ray flare from the blazar
TXS 0506+056~~\cite{IceCube:2018cha} has enabled an analysis of the limits on the velocity
difference between neutrinos and photons, $\delta_{\nu \gamma}$~\cite{Laha:2018hsh,Ellis:2018ogq}.
The result from Ref.~\refcite{Laha:2018hsh} gives
\begin{equation}
\delta_{\nu \gamma} \le 4.2 \times 10^{-12} {{\Delta t}\over{{\rm 7 \ days}}}
\end{equation}
assuming that the time difference is $\sim$7 days.

\section{LIV in the neutrino sector II - Lepton Pair Emission {\it in vacuo}}
\label{vpe}

Since the Lorentz violating operators change the free field behavior and dispersion relation, interactions such as fermion-antifermion pair emission by slightly superluminal neutrinos become kinematically allowed~\cite{Coleman:1998ti,Cohen:2011hx} and can thus cause significant observational effects. 

The dominant pair emission reactions that become allowed if Lorentz invariance is violated are vacuum electron-positron pair emission (VPE) $\nu_i \rightarrow \nu_i + e^+ + e^-$, and its close cousin, neutrino splitting, $\nu_e \rightarrow \nu_e + \nu_i + \bar{\nu_i}$ (where $i$ is a flavor index). These are the reactions involving the leptons with the lightest final state masses. Neutrino splitting can be represented as a rotation of the Feynman diagram for neutrino-neutrino scattering.

\subsection{Lepton pair emission in the [d] = 4 case}
\label{[d]4}

\begin{figure}
\includegraphics[scale=0.55]{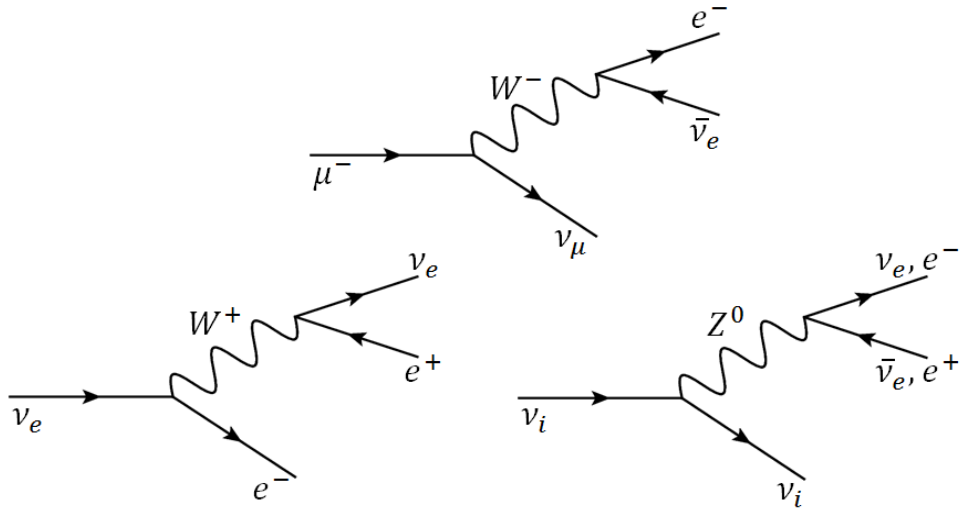}
\caption{Diagrams for muon decay (top), charged current mediated VPE (bottom left), and neutral current mediated neutrino splitting/VPE (bottom right).  Time runs from left to right and the flavor index $i$ represents $e,\mu$, or $\tau$ neutrinos.}\label{fig:Diagrams}
\end{figure}

In this section we consider the constraints on the LIV parameter $\delta_{\nu e}$. We first relate the rates for superluminal neutrinos with that of a more familiar tree level, weak force mediated standard model decay process: muon decay, $\mu^-\rightarrow \nu_\mu + \bar{\nu}_e + e^-$, as the process are very similar (see Figure~\ref{fig:Diagrams}).

For muons with a Lorentz factor $\gamma_{\mu}$ in the observer's frame the decay rate is found to be~\cite{Fermi:1933jpa}
\begin{equation}
{ \Gamma \  =   \gamma_{\mu}^{-1}} {{G_F^2 m_{\mu}^5}\over{192\pi^3}}
\label{mu}
\end{equation}
where $G_F^2 = g^4/(32M_{W}^4)$, is the square of the Fermi constant equal to $1.360 \times 10^{-10} \ {\rm GeV}^{-4}$, with $g$ being the weak coupling constant and $M_{W}$ being the $W$-boson mass in electroweak theory.

We apply the effective energy-dependent mass-squared formalism given by equation (\ref{effectivemass}) to determine the scaling of the emission rate with the $\delta$ parameter and with energy. Noting that for any reasonable neutrino mass, $m_{\nu} \ll 2\delta_{\nu e} E_{\nu}^2$, it follows that $\tilde{m_{\nu}}^2(E) \simeq 2\delta_{\nu e} E_{\nu}^2$.\footnote{At relativistic energies, assuming that the Lorentz violating terms yield small corrections to $E$ and $p$, it follows that $E \simeq p$.}

We therefore make the substitution
\begin{equation}
m_{\mu}^2 \ \rightarrow \ \tilde{m_{\nu}}^2(E) \simeq 2\delta_{\nu e} E_{\nu}^2 
\end{equation}
\noindent from which it follows that
\begin{equation}
{\gamma_{\mu}^2} \ \rightarrow \ {{E_{\nu}^2}\over{2\delta_{\nu e} E_{\nu}^2}} = (2\delta_{\nu e})^{-1}.
\end{equation}
The rate for the vacuum pair emission processes (VPE) is then
\begin{equation}
\Gamma \ \propto \ (2\delta_{\nu e})^{1/2} G_F^2 (2\delta_{\nu e} E_{\nu}^2)^{5/2}  
\label{dimension}
\end{equation}
\noindent which gives the proportionality
\begin{equation}
\Gamma \ \propto \ G_F^2 \ \delta_{\nu e}^3 E_{\nu}^5 
\label{propto}
\end{equation}
\noindent showing the strong dependence of the decay rate on both $\delta_{\nu e}$ and $E_{\nu}$.

The energy threshold for $e^+e^-$ pair production is given by\cite{Stecker:2001vb}
\begin{equation}
E_{th} = m_e\sqrt{{{2}\over{{\delta_{\nu e}}}}} 
\label{threshold}
\end{equation}
We now define $\delta \equiv \delta_{\nu e}$ as given by equation (\ref{sub}). The rate for the VPE process, $\nu \to \nu \,e^+\, e^-$ via the neutral current $Z$-exchange channel, has been calculated to be~\cite{Cohen:2011hx}\footnote{For a different treatment giving compatible results see~\cite{Bezrukov:2011qn}} 
\begin{equation}
\Gamma = \frac{1}{14}\frac{G_F^2 (2\delta)^3E_{\nu}^5}{192\,\pi^3} = 1.31 \times 10^{-14} \delta^3 E_{\rm GeV}^5\ \ {\rm GeV},
\label{G}
\end{equation}
\noindent consistent with our derivation of the dependences on $\delta$ and $E_{\nu}$ given by equation (\ref{propto}).
The mean fractional energy loss per interaction from VPE is 78\%~\cite{Cohen:2011hx}. 
Thus, the energy loss rate is given by ($v_{\nu} \simeq$ 1)
\begin{equation}
\frac{dE}{dx} = -\frac{25}{448} \frac{G_{F}^{2}}{192\pi^3} E^{6}\delta^3
\label{loss}
\end{equation}
It follows from equation (\ref{loss}) that for a superluminal neutrino propagating
from a distance $L$ and observed with a conservatively measured terminal energy $E_{T}$,
the upper limit on $\delta_{\nu}$ is given by~\cite{Cohen:2011hx}
\begin{equation}
E_{T}^{-5} \equiv 0.256 \ \delta_{\nu}^{3} \frac{G_{F}^{2}}{192\pi^{3}}L
\label{ET}
\end{equation}
Using equation (\ref{ET}), the upper limit constraints on $\delta_{\nu}$
from observations of various sources are shown in Table 1.

\begin{table}
\centerline{Table 1: Upper Limits on $\delta_{\nu}$ from equation (\ref{ET})}
\begin{center}
\begin{tabular}{|ccccc|}  \hline \hline\\
Source & Distance & $E_T$ & $\delta_{\nu,max}$  & Reference  \\ 
\hline\\
IceCube upgoing & $>$ 500 km  & 100 TeV  & $1.7 \times 10^{-11}$ & ~\refcite{Cohen:2011hx}  \\
%Crab Nebula & --- & --- & ---- & \refcite{st14} \\ 
TXS 0506+056---& 1.36 Gpc  & 183 TeV & $1.3 \times 10^{-18}$ & ~\refcite{Wang:2020tej} \\

\hline
\end{tabular}
\end{center}
\end{table}

In general, the charged current $W$-exchange channels (CC) contribute as well.  However, this channel is only kinematically relevant for $\nu_e$'s, as the production of $\mu$ or $\tau$ leptons by $\nu_{\mu}$'s or $\nu_{\tau}$'s has a much higher energy threshold due to the larger final state particle masses (equation (\ref{threshold}) with $m_e$ replaced by $m_{\mu}$ or 
$m_{\tau}$), with the neutrino energy loss from VPE being highly threshold dependent. Owing to neutrino oscillations, neutrinos propagating over large distances spend 1/3 of their time in each flavor state. Thus, the flavor population of neutrinos from astrophysical sources is expected to be [$\nu_e$:$\nu_{\mu}$:$\nu_{\tau}$] = [1:1:1] so that CC interactions involving $\nu_e$'s will only be important 1/3 of the time. 

The vacuum \v{C}erenkov emission (VCE) process, $\nu \rightarrow \nu + \gamma$, is also kinematically allowed for superluminal neutrinos. However, since the neutrino has no charge, this process entails the neutral current channel production of a loop consisting of a virtual electron-positron pair followed by its annihilation into a photon. Thus, the rate for VCE is a factor of $\alpha$ lower than that for VPE. 

Neutrino pair emission, a.k.a. neutrino splitting, is unimportant for energy loss of superluminal neutrinos in the $[d] = 4$ case because the fractional energy loss per interaction is very low~\cite{Cohen:2011hx} owing to the small
velocity difference between neutrino flavors obtained from neutrino oscillation data (See Section \ref{osc}). However, as we see in the next section, this is not true in the cases with $[d] > 4$.

\subsection{Vacuum $e^+e^-$ Pair Emission in the $[d] > 4$ cases}.

Using equations (\ref{sub}) and (\ref{prop}) and the dynamical matrix element taken from the simplest case~\cite{Carmona:2012tp}, we can generalize equation (\ref{G}) for arbitrary values of $n = [d] - 4$
\begin{equation}
\Gamma = \frac{G_F^2}{192\,\pi^3}[(1-2s_W^2)^2 + (2s_W^2)^2]\zeta_n \kappa_n^3 \frac{E_\nu^{3n+5}}{M_{Pl}^{3n}} 
\label{G2}
\end{equation} 
\noindent where $s_W$ is the sine of the Weinberg angle ($s_W^2 = 0.231$) and the $\zeta_n$'s are numbers of $\cal{O}$(1). ~\cite{Carmona:2012tp}
 
For the $n = 1$ case we obtain the VPE rate
\begin{equation}
\Gamma = 1.72 \times 10^{-14} \kappa_1^3 E_{\rm GeV}^5\ (E/M_{Pl})^3 \ {\rm GeV},
\label{Gn1}
\end{equation}
and for the $n = 2$ case we obtain the VPE rate
\begin{equation}
\Gamma = 1.91 \times 10^{-14} \kappa_2^3 E_{\rm GeV}^5\ (E/M_{Pl})^6 \ {\rm GeV}.
\label{Gn2}
\end{equation}
 
\section{LIV in the neutrino sector III - decay by neutrino pair emission (neutrino splitting)}
\label{3n} 

The process of neutrino splitting in the case of superluminal neutrinos, i.e., $\nu \rightarrow 3\nu$ is relatively unimportant in the $[d] = 4, n = 0$ owing to the small
velocity difference between neutrino flavors obtained from neutrino oscillation data (see Section \ref{osc}). However, this is not true in the cases with $[d] > 4$. In the presence of $[d] > 4$ $(n > 0)$ terms in a Planck-mass suppressed EFT, the velocity differences between the neutrinos, being energy dependent, become significant~\cite{Maccione:2011fr}. The daughter neutrinos travel with a smaller velocity. The velocity dependent energy of the parent neutrino is therefore greater than that of the daughter neutrinos.  Thus, the neutrino splitting becomes kinematically allowed. Let us then consider the $n = 1$ and $n = 2$ scenarios. 

Neutrino splitting is a neutral current (NC) interaction that can occur for all 3 neutrino flavors. The total neutrino splitting
rate obtained is therefore three times that of the NC mediated VPE process above threshold. Assuming the three daughter neutrinos
each carry off approximately 1/3 of the energy of the incoming neutrino, then 
for the $n = 1$ case one obtains the neutrino splitting rate~\cite{Stecker:2014oxa}
\begin{equation}
\Gamma = 5.16 \times 10^{-14} \kappa_1^3 E_{\rm GeV}^5\ (E/M_{Pl})^3 \ {\rm GeV},
\label{Gsplitn1}
\end{equation}
and for the $n = 2$ case we obtain the neutrino splitting rate
\begin{equation}
\Gamma = 5.73 \times 10^{-14} \kappa_2^3 E_{\rm GeV}^5\ (E/M_{Pl})^6 \ {\rm GeV}.
\label{Gsplitn2}
\end{equation}

The threshold energy for neutrino splitting is proportional to the neutrino mass so that it is negligible compared to that given by equation ({\ref{threshold}). (See also the detailed treatment of neutrino splitting given in Ref.~\cite{Somogyi:2019yis}.)

\section{The IceCube Observations}
\label{ice}
The IceCube collaboration has identified hundreds of events from neutrinos of astrophysical origin with energies above 10 TeV (see Chapter 5 by Halzen and Kheirandish). 

There are are four indications that the the bulk of cosmic neutrinos observed by IceCube with energies above 0.1 PeV are of extragalactic origin: 

(1) The arrival distribution of the reported events with $E > 0.1$ PeV observed by IceCube above atmospheric background is roughly consistent with isotropy~\cite{IceCube:2016tpw}.

(2) The diffuse galactic neutrino flux is expected to be well below that observed by IceCube~\cite{Stecker:1978ah,IceCube:2016tpw,Denton:2017csz}.

(3) Upper limits on diffuse galactic $\gamma$-rays in the TeV-PeV energy range imply that galactic neutrinos cannot account for the neutrino flux observed by IceCube~\cite{Ahlers:2013xia}.

Above 60 TeV, the IceCube data are crudely consistent with a spectrum given by $E_{\nu}^2(dN_{\nu}/dE_{\nu}) \simeq \ 10^{-8}$ \ ${\rm GeV}{\rm cm}^{-2}{\rm s^{-1}}$ extending up to an energy $\sim$ 2.2 PeV, but dropping off above that energy~\cite{IceCube:2020acn}. 
 
IceCube has also reported a candidate neutrino induced event at the Glashow resonance of 6.3 PeV~\cite{IceCube:2021rpz}. At this energy, electrons in the IceCube volume provide enhanced target cross sections for $\bar{\nu}_{e}$'s through the $W^-$ resonance channel, $\bar{\nu}_{e} + e^- \rightarrow W^- \rightarrow ~shower$ at the resonance energy $E_{\bar{\nu}_{e}} = M_W^2/2m_{e} = 6.3$ PeV~\cite{Glashow:1960zz}.

\section{The Energy Spectrum from Extragalactic Superluminal Neutrino Propagation}
\label{prop}

Monte Carlo techniques can be used to determine the effect of neutrino splitting and VPE on putative superluminal neutrinos propagating from cosmological distances under the assumption of the dominance of Planck mass suppressed LIV operators with $[d] > 4$~\cite{Stecker:2014oxa}. The Monte Carlo codes used in Ref.~\refcite{Stecker:2014oxa} take account of energy losses by both neutrino splitting and VPE as well as redshifting of neutrinos emitted from sources at cosmological distances. As in Ref.~\refcite{Stecker:2014xja} and Ref.~\refcite{Stecker:2014oxa}, we here consider a scenario where the neutrino sources have a redshift distribution that follows that of the star formation rate and further assume a simple source spectrum roughly proportional to $E^{-2}$ between 100 TeV and 100 PeV. The redshift distribution describing star formation appears to be roughly applicable for both active galactic nuclei and $\gamma$-ray bursts. 

The Monte Carlo events were generated using these two distributions. The final results were normalized to an energy flux of $E_{\nu}^2(dN_{\nu}/dE_{\nu}) \simeq 10^{-8}\  {\rm GeV}{\rm cm}^{-2}{\rm s}^{-1}{\rm sr}^{-1}$, as is consistent with the IceCube data for both the southern and northern hemisphere for energies between 60 TeV and 2 PeV~\cite{IceCube:2020acn}. 

The Monte Carlo runs in Ref.~\refcite{Stecker:2014oxa} were made using threshold energies between 10 PeV and 40 PeV for the VPE process, corresponding to values of $\delta_{\nu e}$ between $5.2\times10^{-21} \ {\rm and}~ 3.3 \times 10^{-22}$. By propagating the test neutrinos including energy losses from VPE, neutrino splitting, and redshifting using this code, final neutrino spectra were obtained and compared with the IceCube results.

Given that neutrinos detected by IceCube are extragalactic, cosmological effects should be taken into account in deriving new LIV constraints. The reasons are straightforward. As opposed to the extinction of high energy extragalactic photons through electromagnetic interactions~\cite{Stecker:1992wi}, neutrinos survive from all redshifts because they only interact weakly.  It follows that since the universe is transparent to neutrinos, most of the cosmic PeV neutrinos will come from sources at redshifts between $\sim$0.5 and $\sim$2 where most of the energy in the universe from astrophysical sources is produced. Therefore, along with energy losses by VPE~\cite{Cohen:2011hx} and neutrino splitting, energy losses by redshifting of neutrinos and the effect of the cosmological $\Lambda$CDM redshift-distance relation 
\beq
D = {{c}\over{H_0}}\int\limits_0^z\frac{dz'}{(1+z')\sqrt{\Omega_{\rm\Lambda} + \Omega_{\rm M} (1 + z')^3}}
\eeq 
need to be included in the determination of $\delta_{\nu}$. 

As in Refs.~\refcite{Stecker:2014xja}~\refcite{Stecker:2014oxa}, we assume here a flat $\Lambda$CDM universe with a Hubble constant of $H_0 =$ 67.8 km s$^{-1}$ Mpc$^{-1}$ along with $\Omega_{\rm\Lambda}$ = 0.7 and $\Omega_{\rm M}$ = 0.3.  Thus the energy loss due to redshifting is given by
\begin{equation}
-(\partial \log  E/\partial t)_{redshift} =  H_{0}\sqrt{\Omega_{m}(1+z)^3 +
  \Omega_{\Lambda}}.
\label{redshift}  
\end{equation}
The decay widths for the VPE process are given by equations (\ref{Gn1}) and (\ref{Gn2}) for the cases $n = 1$ and $n = 2$ respectively while those for neutrino splitting are given by equations (\ref{Gsplitn1}) and (\ref{Gsplitn2}).

\subsection{[d] = 4 $\cal{CPT}$ Conserving Operator Dominance}

In their seminal paper, using equation (\ref{G}), Cohen and Glashow~\cite{Cohen:2011hx}} showed how the VPE process in the $[d] = 4$ case implied powerful constraints on LIV. They obtained an upper limit of $\delta < \cal{O}$ $(10^{-11})$ based on the initial observation of high energy neutrinos by IceCube~\cite{IceCube:2011mzm}. 
Based on later IceCube observations, using the results of Ref.~\refcite{Cohen:2011hx} and later IceCube results~\cite{Aartsen:2013jdh} and assuming a distance to the neutrino source of 1 Gpc, an upper limit of $\delta < \cal{O}$ $(10^{-18})$ was obtained in Ref.~\refcite{Borriello:2013ala}. 

General predictions of limits on $\delta$ with cosmological factors taken into account were then made, with the predicted spectra showing a pileup followed by a cutoff~\cite{Stecker:2014oxa}. Using the energy loss rate given by equation (\ref{G}), a value for $\delta_{\nu e} < 10^{-20}$ was obtained based on a model of redshift evolution of neutrino sources based on the and using Monte Carlo techniques to take account of propagation effects as discussed in Section \ref{prop}~\cite{Stecker:2014xja,Stecker:2014oxa}. The upper limit on $\delta_{e}$ is given by $\delta_{e} \le 5 \times 10^{-21}$. Taking this into account, one gets the constraint $\delta_{\nu} \le (0.5 - 1) \times 10^{-20}$. The spectra derived therein for the $[d] = 4$ case also showed a pileup followed by a cutoff. The predicted cutoff is determined by redshifting the threshold energy effect. Monte Carlo techniques to take account of propagation effects as discussed in Section \ref{prop}~\cite{Stecker:2014xja,Stecker:2014oxa}. The upper limit on $\delta_{e}$ is found to be $\delta_{e} \le 5 \times 10^{-21}$. Taking this into account, one gets the constraint $\delta_{\nu} \le (0.5 - 1) \times 10^{-20}$. The spectra derived therein for the $[d] = 4$ case show a slight pileup followed by a cutoff. The predicted cutoff is determined by redshifting the threshold energy effect. The blue curve in Figure \ref{neutrinospec} holds for the [d] = 4 case because the energy losses are only from VPE .

\subsection{[d] = 6 $\cal{CPT}$ Conserving Operator Dominance}
\label{d6} 

In both the $[d] = 4$ and $[d] = 6$ cases, the best fit matching the theoretical propagated neutrino spectrum, normalized to an energy flux of $E_{\nu}^2(dN_{\nu}/dE_{\nu}) ~\simeq 10^{-8} {\rm GeV}{\rm cm}^{-2}{\rm s}^{-1}{\rm sr}^{-1}$ below $\sim$~0.3 PeV, is assumed in the calculations.

\begin{figure}[!t]
{\includegraphics[scale=0.8]{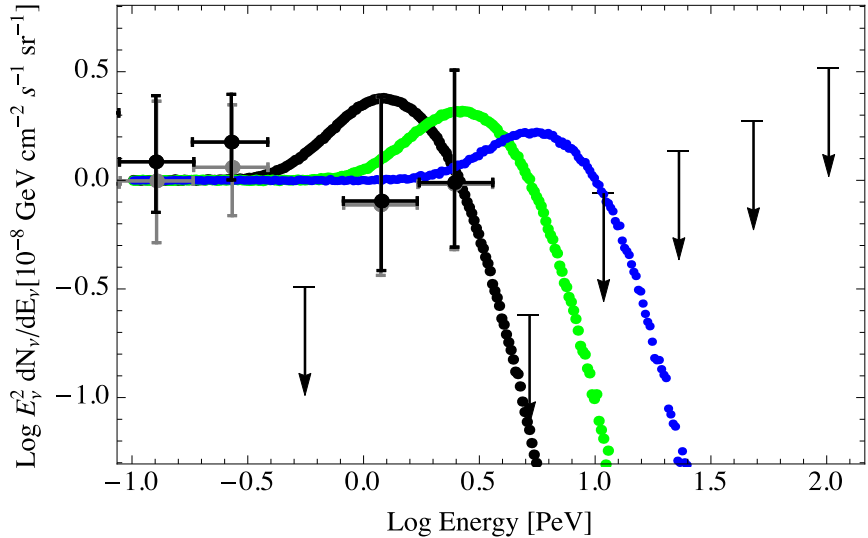}}
\caption{Calculated n = 2 spectra taking into account of all three processes
(redshifting, neutrino splitting, and VPE) occurring simultaneously for rest frame VPE threshold energies  of 10 PeV (black), 20 PeV (green), and 40 PeV (blue)~\cite{Stecker:2014oxa}. The six year IceCube data
are also shown (see Chapter 5 by Halzen and Kheirandish.)}
\label{thresholdeffects}
\end{figure}

As found before~\cite{Stecker:2014xja,Stecker:2014oxa}, the best fit match to the IceCube data corresponds to a VPE rest-frame threshold energy $E_{\nu, \rm th} = 10$ PeV, as shown in Figure~\ref{thresholdeffects}. Such a threshold energy corresponds to a value for $\delta_{\nu e} \equiv \delta_{\nu} - \delta_e \le \ 5.2 \times 10^{-21}$. Given that $\delta_e \le \ 5 \times 10^{-21}$, it is again found that $\delta_{\nu} \le (0.5 - 1) \times 10^{-20}$\footnote{We note that one can not assume that $\delta_{\nu}$ and $\delta_e$ are equal. Models can be constructed where $\delta_{\nu}$ and $\delta_e$ are independent and it has even been suggested that LIV may occur only in the neutrino sector~\cite{Carmona:2012tp}.}$^{,}$  
\footnote{The $\nu$ is used here generically for all three neutrino flavors, $\nu_e, \nu_\mu$, and $\nu_{\tau}$ and $\delta_{\nu}$ is for assumed for all three flavors as their velocity differences are very small (see Sect. \ref{osc})}.

As indicated in Figure \ref{thresholdeffects}, values of $E_{\nu, \rm th}$ less than 10 PeV are inconsistent with the IceCube data. The result for a 10 PeV rest-frame threshold energy is just consistent with the IceCube results, giving a cutoff effect above 2 PeV. The apparent cutoff in the IceCube observations indicated in Figure~\ref{thresholdeffects} led to the suggestion of LIV as a possible explanation for the lack of observed neutrinos above 2 PeV~\cite{Stecker:2014xja,Stecker:2014oxa}. However, should confirmed events be found corresponding to neutrino energies above 2 PeV, the values for $\delta_{\nu}$ given above would be upper limits.

In the case of the $\cal{CPT}$ conserving $[d] = 6$ operator (n = 2) dominance, as in the $[d] = 4$ case, the results shown in Figure~\ref{thresholdeffects} show a predicted high-energy drop off in the propagated neutrino spectrum near the redshifted VPE threshold energy and a pileup in the spectrum below that energy. This pileup is caused by the propagation of higher energy neutrinos in energy space down to energies within a factor of $\sim$ 5 below the VPE threshold. 

\begin{figure}[!t]
{\includegraphics[scale=1.21]{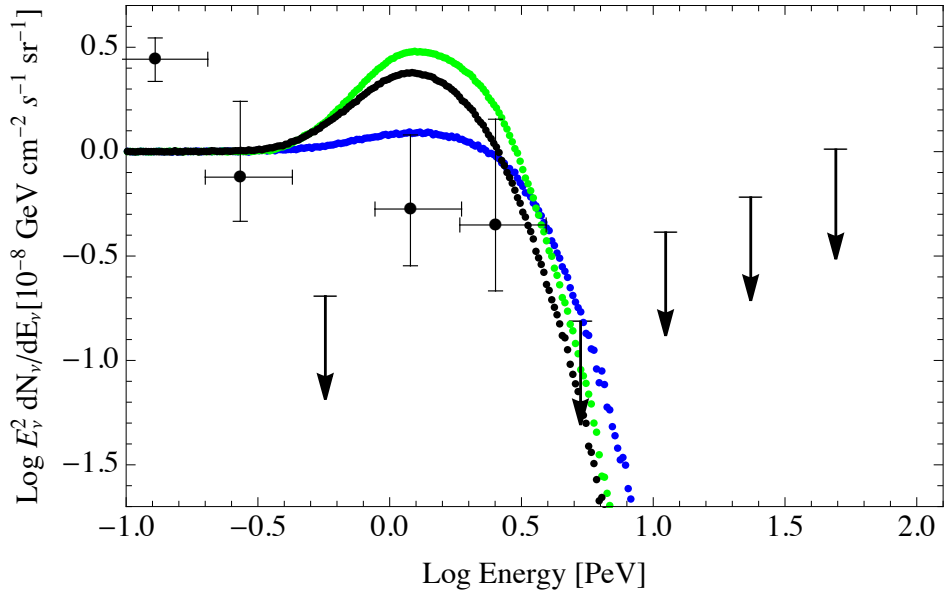}}
\caption{Separately calculated n = 2 neutrino spectra with the VPE case shown in blue and the neutrino splitting case shown in green. The black spectrum takes account of all three processes
(redshifting, neutrino splitting, and VPE) occurring simultaneously. The rates for all cases are fixed by setting the rest frame threshold energy for VPE at 10 PeV~\cite{Stecker:2014oxa}. The neutrino spectra are normalized to the six year IceCube data.}
\label{neutrinospec}
\end{figure}

As the pileup effect shown in Figure \ref{neutrinospec}, that is caused by the neutrino splitting process is more pronounced than that caused by the VPE process. This is because neutrino splitting produces two new lower energy neutrinos per interaction. This difference would be a potential way of distinguishing a dominance of $[d] > 4$ Planck-mass suppressed interactions from $[d] = 4$ interactions. Thus, with better statistics in the energy range above 100 TeV,a significant pileup effect would be a signal of Planck-scale physics. Pileup features are indicative of the fact that fractional energy loss from the last allowed neutrino decay before the VPE process ceases is 78\%~\cite{Cohen:2011hx} and that for neutrino splitting is taken to be 1/3. The pileup effect is similar to that of energy propagation for ultrahigh energy protons near the GZK threshold~\cite{Stecker:1989ti}.

Values of $E_{\nu, \rm th}$ less than 10 PeV are inconsistent with the IceCube data. The result for a 10 PeV rest-frame threshold energy, corresponding to $\delta_{\nu e} = \ 5.2 \times 10^{-21}$, is just consistent with the IceCube results, giving a cutoff effect above $\sim$2 PeV. Thus for the conservative case of no-LIV effect, {\it e.g.}, if one assumes a cutoff in the intrinsic neutrino spectrum of the sources, or one assumes a slightly steeper PeV-range neutrino spectrum proportional to $E_{\nu}^{-2.3}$, we previously obtained the constraint on superluminal neutrino velocity, $\delta_{\nu} = \delta_{\nu e} + \delta_e  \le \ 1.0 \times 10^{-20}$~\cite{Stecker:2014xja}.

In the case of the $\cal{CPT}$ conserving [d] = 6 operator (n = 2) dominance, the results in Figure~\ref{thresholdeffects} appear to show a high-energy drop off in the propagated neutrino spectrum near the redshifted VPE threshold energy and may be consistent with a pileup in the spectrum below that energy, given the statistics of small numbers. This predicted drop off may be a possible explanation for the lack of observed neutrinos above 2 PeV~\cite{Stecker:2014xja,Stecker:2014oxa}.\footnote{The report of a possible Glashow resonance from a 6.3 PeV $\bar{\nu_{e}}$ may tighten the above constraints on $\delta_{\nu e}$.} A pileup would be caused by the propagation of the higher energy neutrinos in energy space down to energies within a factor of $\sim$5 below the VPE threshold. This is indicative of the fact that fractional energy loss from the last allowed neutrino decay before the VPE process ceases is 0.78~\cite{Cohen:2011hx} and that for neutrino splitting is taken to be 1/3. The pileup effect is similar to that of energy propagation for ultrahigh energy protons near the GZK threshold~\cite{Stecker:1989ti}. As is shown
in Figure \ref{neutrinospec}, including neutrino splitting in the calculation increases the pileup effect.

\begin{figure}[!t]
{\includegraphics[scale=0.85]{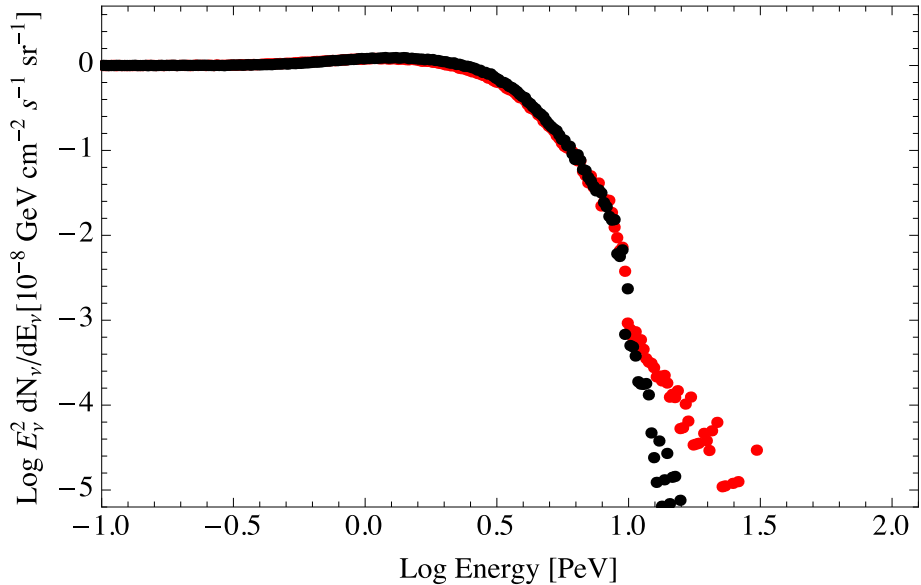}}
\caption{Calculated n = 0  (red) and n = 2 (black) neutrino spectrum obtained for the VPE process only (no neutrino splitting) simultaneously with redshifting. The rates for all cases are fixed by setting the threshold energy for VPE at 10 PeV~\cite{Stecker:2014oxa}.}
\label{vpeonly}
\end{figure}

\begin{figure}[!t]
{\includegraphics[scale=1.0]{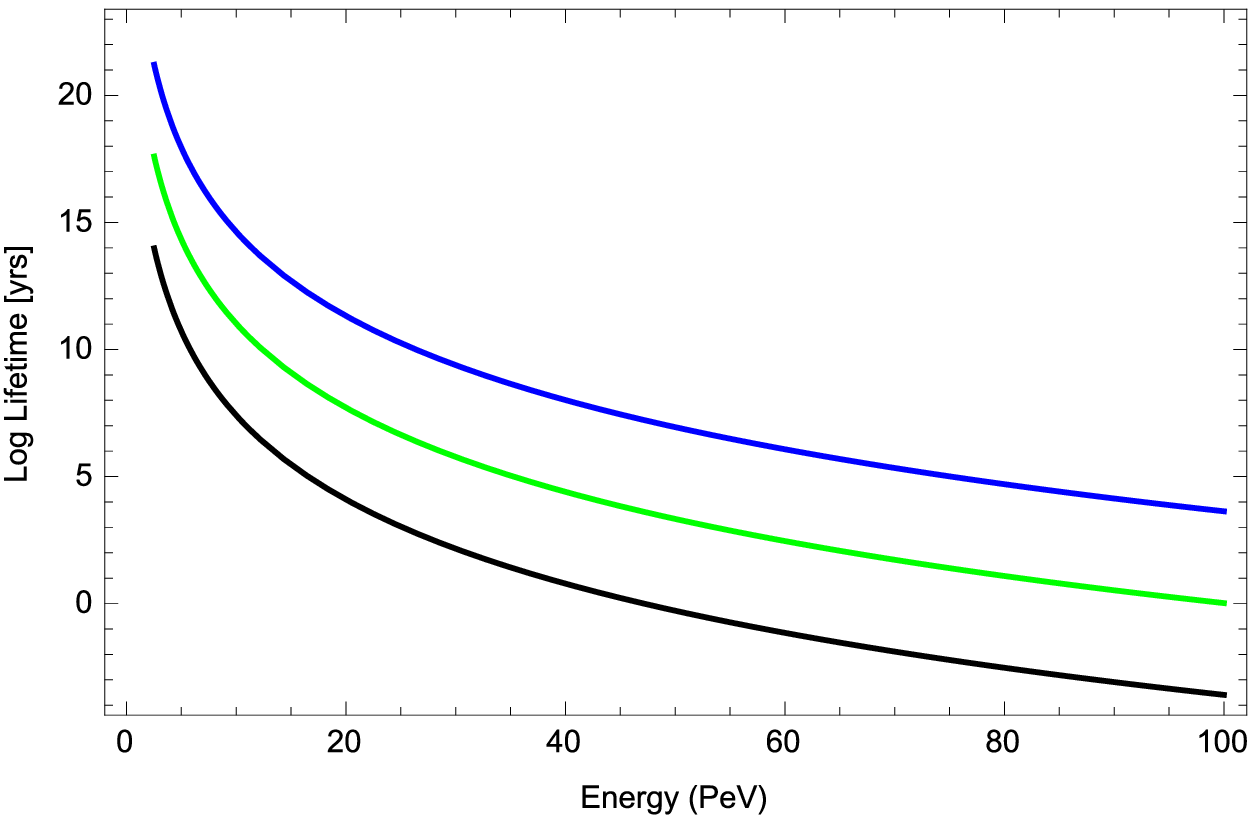}}
\caption{Mean decay times for neutrino splitting process in the n = 2 case obtained by setting the threshold energy for VPE at 10 PeV (black), 20 PeV (green), and 40 PeV (blue)~\cite{Stecker:2014oxa}.}
\label{lifetimes}
\end{figure}

In order to determine threshold effects for the VPE process, a Monte Carlo routine was employed in Ref.~\refcite{Stecker:2014oxa} to find the opening up of phase space, assuming the same LIV parameters for every particle but with an electron mass for two of the outgoing states.  It was found that the entirety of phase space is available when the energy reaches about 1.6 times that of threshold.  Threshold effects should therefore have little impact on the results of the Monte Carlo calculation, as above 1.6$E_{th}$ the full phase space is open.  A Monte Carlo exploration of phase space for neutrino splitting yields similar results. However, the threshold for this reaction is in the GeV range meaning that full rates apply throughout our calculation. 

In practice, neutrinos near threshold rarely pair produce before redshifting below their threshold energy since their mean decay times increase with their decreasing energy as they propagate. This also justifies the assumption that the neutrino splitting and VPE rates are similar per decay channel.    

Throughout the above calculation it was  assumed that a neutrino loses 0.78 of its initial energy per VPE interaction.  Equation (\ref{G2}) shows that the VPE rates 
do not differ by more than 45\% between the $n = 0$ and 
$n = 2$ cases. This reflects the difference in the phase space 
factors, since the dynamical matrix elements are the same, indicating that this is also the maximum deviation in the fraction of energy carried off by the neutrino 
in VPE. It is likely that the deviation would be at most a third of that in a
three-body decay, {\it viz.}, 15\% meaning that the resulting energy fraction for the 
$n = 2$ case could be as high as 0.25. This effect was found to produce no discernible difference in the spectra.  In Ref.~\refcite{Stecker:2014oxa} we also tested an energy fraction of 0.5 and found that even this extreme case would generate no observational consequences on the pileup effect. 

In Figure~\ref{vpeonly}, plots the VPE process alone (along with redshifting) for the $\cal{CPT}$- conserving cases $n = 0$ and $n = 2$.  It can be seen that the resulting spectra are indistinguishable below threshold.  Events above the redshifted threshold pair-produce in relatively short times compared to cosmological timescales regardless of the energy dependence, making the spectra for $n = 0$ and $n = 2$ below the redshifted threshold indistinguishable.  One can only see the expected differences in the steepening of the spectra for energies above threshold owing to the rate differences between
$n = 0$ and $n = 2$ given by equation (\ref{G2}).

As can be seen in Figure ~\ref{lifetimes}, the mean decay times increase for the neutrino splitting process with increasing choice of VPE threshold.  The increased mean decay times have the effect of reducing the pileup for increased choice of threshold as fewer neutrino splitting events will occur.  Thus the pileup becomes a somewhat less sensitive test of Planck-scale effects with increasing threshold energies. Figure~\ref{thresholdeffects} shows the effects of choosing different threshold energies. The dominant process continues to be that of neutrino splitting but with decreasing importance.

\subsection{[d] = 5 $\cal{CPT}$ Violating Operator Dominance}
\label{CPTodd}

In the n=1 case, the dominant [d] = 5 operator violates $\cal{CPT}$.

Both VPE and neutrino splitting generate
a particle-antiparticle lepton pair. The particles
of this pair will have opposite helicities. This holds true
for both Dirac neutrinos and Majorana neutrinos, with the
later being their own antiparticles, In this
case, one of the pair particles will be
superluminal ($\delta > 0$) whereas the other particle will be subluminal 
($\delta < 0$)~\cite{Pospelov:2012zva}. Thus, of the daughter particles,
one will be superluminal and interact, while the other will only redshift.
The overall result in the $[d] = 5$ case is that no strong spectral cutoff occurs~\cite{Stecker:2014xja}.

However, the IceCube detector cannot distinguish neutrinos from antineutrinos. 
The incoming $\nu (\bar{\nu}$) generates a shower in the detector, allowing
a measurement of its energy and direction. Even in cases where there is a muon
track, the charge of the muon is not determined. 

There would be an exception for electron antineutrinos at 6.3 PeV, 
given an expected enhancement in the event rate at the $W^{-}$ Glashow 
resonance, since this resonance only occurs with $\bar{\nu_{e}}$. 
candidate Glashow resonance event has now been reported by the IceCube collaboration~\cite{Glashow:1960zz}. 
A confirmation of such resonance events might favor
a $\cal{CPT}$-odd interpretation, particularly if they are rarer than expected.

\begin{figure}[!t]
{\includegraphics[scale=1.22]{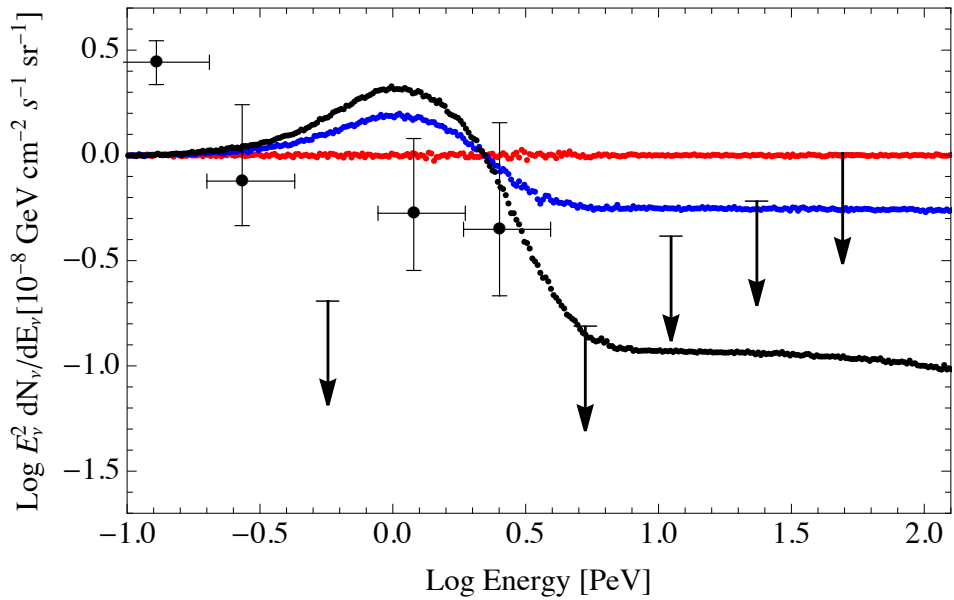}}
\caption{Calculated n = 1 neutrino spectra assuming 100\% (black), 50\% (blue) and 0\% (red) initial superluminal neutrinos (antineutrinos). The neutrino spectra are normalized to the IceCube
data.}
\label{CPTodd}
\end{figure}

Figure~\ref{CPTodd} shows the results in the $\cal{CPT}$-violating $n = 1$ case, assuming 100\%, 50\% and 0\% initial superluminal neutrinos (antineutrinos) and propagating the spectrum using a
 Monte Carlo program and taking account of the fact that in all cases, one of the daughter leptons is subluminal and therefore does not undergo further interactions. As a sanity check, it was found that in the 0\% case only redshifting occurs, preserving the initial $E^{-2}$ spectrum. The other cases show the effect of VPE and neutrino splitting by both the initial fraction of superluminal neutrinos and the superluminal daughter neutrinos.

Thus, as opposed to the
$\cal{CPT}$-conserving $n = 2$ case,  no clearly observable cut off is produced, with the possible unrealistic exception of postulating that only superluminal $\nu$'s (or superluminal $\bar{\nu}$'s) are produced in cosmic sources. That case, shown in black in Fig.~\ref{CPTodd}, as well as the other case of postulating no initial superluminal neutrinos, shown in red, are shown for illustrative purposes. The 50/50 case, shown in blue, is more realistic. 

In the n = 1, $\cal{CPT}$-odd case, the
details of the kinematics are different from the n = 0 and n = 2, $\cal{CPT}$-even
cases because in the $\cal{CPT}$-odd case the signs of $\delta$ are
opposite for $\nu$'s and $\bar{\nu}$'s. If we assume that they are equal and opposite, then the rate given in equation (\ref{G2}) would maximally be altered by replacing the $\delta$ with 2$\delta$.  Since the source kinematics dominate as the daughter energies are comparable, doubling delta should overestimate their contribution to the overall rate. Applying Monte Carlo techniques to explore the phase space~\cite{Stecker:2014oxa}, one finds that the subliminal particle will carry away a slightly higher fraction of the energy after the split ($\sim 40$\%) in the $\cal{CPT}$-odd case. By making these modifications to the Monte Carlo code one finds that there is little observational difference between the modified results and those obtained assuming the same rate as given by equation (\ref{G2}).  An exact treatment of the kinematics for $\cal{CPT}$-odd, which are complex, are therefore unnecessary and our spectral results in the $\cal{CPT}$-odd case given in Figure \ref{CPTodd} are a good approximation to an exact treatment.

\section{Summary of the results from LIV kinematic effects for superluminal neutrinos}
\label{sum}

In the previous sections we have explored and summarized the effects of $[d] > 4$ Planck-mass suppressed operators on the propagation and resulting energy spectrum of superluminal neutrinos of extragalactic origin. We have expressed these Lorentz violating perturbations as a modifications of the energy-momentum dispersion relation in the form  $\delta_{\nu e} \simeq \delta_{\nu} \equiv \delta_n$ (see discussion in IIIA) for Planck mass suppressed energy dependent values of $\delta_{n}$ as defined in equation (\ref{sub}). These terms can arise from higher dimension operators in the EFT formalism~\cite{Colladay:1998fq}.

In the SME EFT formalism~\cite{Kostelecky:1988zi,Colladay:1998fq,Kostelecky:2009zp}, if the apparent cutoff above $\sim 2$ PeV in the neutrino spectrum shown in Figure \ref{thresholdeffects} is caused by LIV, this would result from an EFT with either a dominant $[d] = 4$ term with $\mathaccent'27 c^{(4)} = -\delta_{\nu e} = 5.2 \times 10^{-21}$, or by a dominant $[d] = 6$ term with $\mathaccent'27 c^{(6)} = -\kappa_2/M_{Pl}^2 \ge - 5.2 \times 10^{-35}$ GeV$^{-2}$ (Refs.~\refcite{Kostelecky:2011gq,Stecker:2014oxa}, see appendix B)
\footnote{The IceCube collaboration has looked for deviations in $\nu_{\mu}$ oscillations between horizontal and vertical muon fluxes of atmospheric origin in the detector~\cite{OSC} (See Section 4). The horizontal path length is much shorter than the vertical path length and is used for normalization. They obtain constraints of order $10^{-36}$ GeV$^{-2}$ on the $[d] = 6$  operator involved.} 
Such a cutoff would not occur if the dominant LIV term is a $\cal{CPT}$-violating the $[d] = 5$ operator. Future detections of
astrophysical neutrinos with energy above $\sim$2 PeV would indicate that the above numbers should be considered to be upper limits on these parameters.%\footnote{The neglected contribution from CC interactions is flavor suppressed by a factor of three relative to the NC channel.}  \footnote{There is further flavor suppression of VPE relative to neutrino splitting, as neutrino splitting involves three possible final state neutrino flavor decay channels with negligible velocity differences, whereas VPE involves only the electronic sector. Hence, neutrino splitting becomes the dominant energy loss mechanism and the charged current contribution to the VPE rate is a small correction to the overall observational signal. This correction will not affect the cutoff energy, but will only produce an unobservable contribution to the pileup effect.}

Furthermore, we note that, should the $\cal{CPT}$-violating [d] = 5 operator dominate, we would find an absence of a clear cutoff in the propagated neutrino spectrum. 

\section{LIV in the Ultrahigh Energy Cosmic Ray Spectrum and the Subsequent Ultrahigh Energy Neutrino Spectrum}

\subsection{The GZK Effect}

Shortly after the discovery  of the 3K cosmic background radiation
(CBR),  Greisen  \cite{Greisen66}  and  Zatsepin  and  Kuz'min  \cite{ZatsKuz66}
predicted that pion-producing interactions  of such cosmic ray protons
with the CBR should produce a spectral cutoff at $E \sim$ 50 EeV.  The
flux  of  ultrahigh energy  cosmic  rays  (UHECR)  is expected  to  be
attenuated by such photomeson  producing interactions.  This effect is
generally known as the "GZK  effect".  Owing to this effect, protons
with energies above $\sim$100~EeV  should be attenuated from distances
beyond $\sim 100$ Mpc because  they interact with the CBR photons with
a resonant photoproduction of pions \cite{Stecker:1968uc}. The  GZK  effect 
is  not  a  true cutoff,  but  a  suppression of  the
ultrahigh  energy  cosmic  ray  flux  owing  to  an  energy  dependent
propagation time  against energy losses  by such interactions,  a time
which is only $\sim$300 Myr  for 100 EeV protons \cite{Stecker:1968uc}.  At high
redshifts, $z$,  the target photon density increases  by $(1+z)^3$ and
both the photon and initial cosmic ray energies increase by $(1+z)$.
If the source  spectrum is  hard enough, there  could also be  a relative
enhancement just  below the  ``GZK energy'' owing  to a  ``pileup'' of
cosmic rays starting out at  higher energies and crowding up in energy
space at or below the predicted GZK cutoff energy~\cite{Stecker:1989ti} (See also Section
\ref{d6}).  

\subsection{The Effect of LIV Kinematics on the GZK Process}

We now consider the kinematics of the photomeson production process leading  
to the GZK effect. Near threshold, where single pion production dominates,

\begin{equation}
p + \gamma \rightarrow p + \pi.
\end{equation}

Using the  normal Lorentz  invariant kinematics, the  energy threshold
for  photomeson interactions  of UHECR  protons of  initial laboratory
energy $E$ with  low energy photons of the  CBR with laboratory energy
$\omega$, is  determined by the relativistic invariance  of the square
of  the  total  four-momentum   of  the  proton-photon  system.   This
relation, together with the threshold inelasticity relation $E_{\pi} =
m/(M  +  m)  E$  for  single pion  production,  yields  the  threshold
conditions for head on collisions in the laboratory frame

\begin{equation}
4\omega E = m(2M + m)
\end{equation}

\noindent for the proton, and

\begin{equation}
4\omega E_{\pi} = {{m^2(2M + m)} \over {M + m}}
\label{pionproduction}
\end{equation}

\noindent in terms of the pion energy, where M is the rest mass of the
proton and m is the rest mass of the pion.%~\cite{st68}.

If LI  is broken so  that $\delta_\pi~ >~  \delta_p$ ($\delta_{\pi p} > 0)$, it follows  from equations
(\ref{LIVdispersion}) and   (\ref{groupvel})   that  the
threshold energy for  photomeson production is altered~\cite{Coleman:1998ti} because the  square of the
four-momentum  is shifted  from  its  LI form  so  that the  threshold
condition in terms of  the pion energy becomes%\footnote{We assume here
%that protons  and pions are kinematically independent  entities. For a
%treatment of these  particles as composites of quarks  and gluons, see
%Ref.~\cite{ga04}.}

\begin{equation}
4\omega E_{\pi}  = {{m^2(2M +  m)} \over {M  + m}} + 2  \delta_{\pi p}
E_{\pi}^2
\label{LIVpi}
\end{equation}

\noindent 

Equation (\ref{LIVpi})  is a quadratic  equation with real  roots only
under the condition

\begin{equation}
\delta_{\pi p}  \le {{2\omega^2(M  + m)} \over  {m^2(2M +  m)}} \simeq
\omega^2/m^2.
\label{root}
\end{equation}

Defining  $\omega_0 \equiv  kT_{CBR} =  2.35 \times  10^{-4}$  eV with
$T_{CBR} = 2.725\pm 0.02$ K, equation (\ref{root}) can be rewritten

\begin{equation}
\delta_{\pi p} \le 3.23 \times 10^{-24} (\omega/\omega_0)^2.
\label{CG}
\end{equation}

If LIV occurs and $\delta_{\pi p} > 0$, photomeson production can only
take place for interactions of  CBR photons with energies large enough
to satisfy equation (\ref{CG}). This condition, together with equation
(\ref{LIVpi}), implies  that while photomeson  interactions leading to
GZK suppression can occur for ``lower energy'' UHE protons interacting
with higher energy CBR photons on the Wien tail of the spectrum, other
interactions involving higher energy  protons and photons with smaller
values  of  $\omega$ will  be  forbidden.   Thus,  the observed  UHECR
spectrum may  exhibit the characteristics of GZK  suppression near the
normal GZK threshold, but the UHECR spectrum can "recover" at higher
energies  owing to  the  possibility that  photomeson interactions  at
higher  proton energies  may be  forbidden.   

The  kinematical  relations   governing  photomeson  interactions  are
changed  in  the  presence  of  even  a  small  violation  of  Lorentz
invariance.      Following     equations   (\ref{LIVdispersion})     and
(\ref{deltadef}), we denote

\begin{equation}
E^2=p^2+2\delta _a p^2 +{m_a}^2
\label{dispersion}
\end{equation}

\noindent where $\delta _a$ is  the difference between the MAV for the
particle {\it a} and the speed  of light in the low momentum limit ($c
= 1$).

The square of the cms energy of particle $a$ is then given by

\begin{equation}
\sqrt{s_a} = \sqrt{E^2-p^2} = \sqrt{2\delta_a p^2 + m_a^2} \ge 0.
\label{restmass}
\end{equation}

Owing to  LIV, in the cms the  particle will not generally  be at rest
when $p = 0$ because

\begin{equation}
v = {{\partial E} \over {\partial p}} \neq {{p}\over {E}}.
\end{equation}

The modified kinematical relations containing LIV have a strong effect
on the amount of energy transferred  from a incoming proton to the pion
produced in  the subsequent interaction, {\it  i.e.}, the inelasticity
\cite{Alfaro:2002ya,Stecker:2009hj}.

\begin{figure}[!t]
{\includegraphics[scale=0.7]{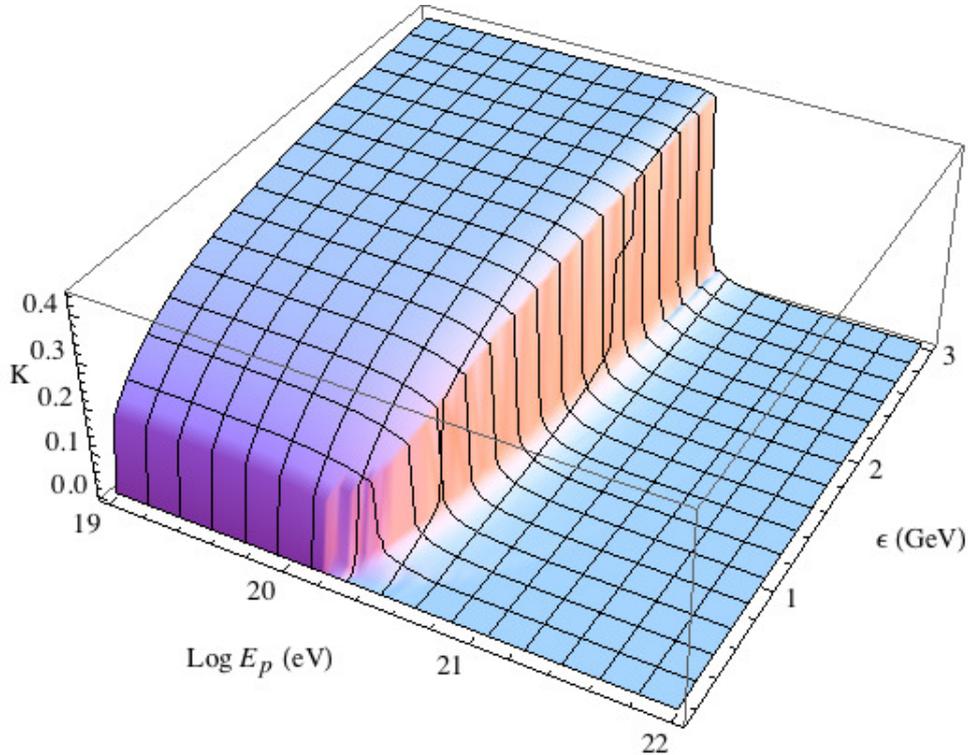}}
%\vspace{-1.5cm}
\caption{The  calculated  proton  inelasticity  modified  by  LIV  for
$\delta_{\pi  p} =  3 \times  10^{-23}$ as  a function  of  CBR photon
energy and proton energy \protect \cite{Stecker:2009hj,Scully:2008jp}.}
\label{inelasticity}
\end{figure}

Figure  \ref{inelasticity} shows  the  calculated proton  inelasticity
modified by LIV for a value of $\delta_{\pi p} = 3 \times 10^{-23}$ as
a function  of both CBR  photon energy and proton  energy \cite{Stecker:2009hj}.
Other choices for $\delta_{\pi p}$ yield similar plots.  The principal
result  of changing the  value of  $\delta_{\pi p}$  is to  change the
energy  at which  LIV effects  become  significant.  For  a choice  of
$\delta_{\pi p}  = 3 \times  10^{-23}$, there is no  observable effect
from LIV for $E_{p}$ less  than $\sim200$ EeV.  Above this energy, the
inelasticity  precipitously drops  as the  LIV term  in the  pion rest
energy approaches $m_{\pi}$.

\begin{figure}[!t]
{\includegraphics[scale=1.15]{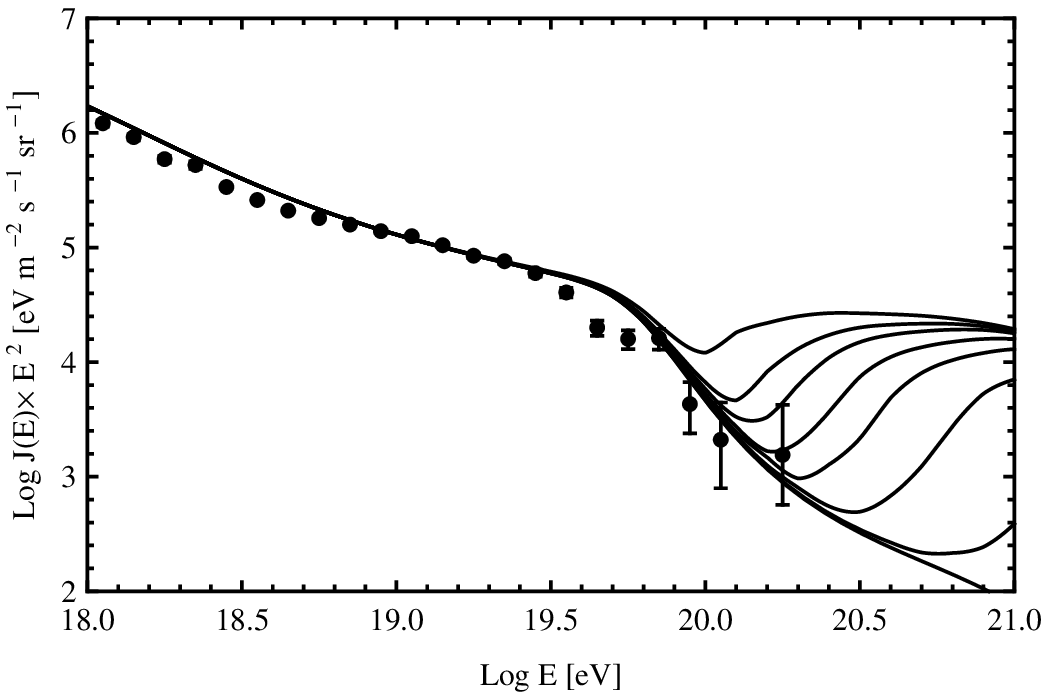}}
%\vspace{-1.5cm}
\caption{Comparison of  the latest Auger data  with calculated spectra
for various  values of  $\delta_{\pi p}$, taking  $\delta_p =  0$ (see
text).  From top to bottom,  the curves give the predicted spectra for
$\delta_{\pi p}  = 1  \times 10^{-22}, 6  \times 10^{-23},  4.5 \times
10^{-23}, 3 \times 10^{-23} ,  2 \times 10^{-23}, 1 \times 10^{-23}, 3
\times 10^{-24}, 0$ (no Lorentz violation) 
\protect \cite{Stecker:2009hj,Scully:2008jp}.}
\label{Auger}
\end{figure}

\subsection{The Photomeson Neutrino Spectrum}
\label{neutrinos}

We now turn our attention to calculating the photomeson neutrino spectrum that would result from the 
UHECR calculation detailed in the previous section. To determine this, we use the data on the cross
section for pion production compiled in reference \cite{Arndt:2002xv} and summarized in reference
\cite{Bernstein:2007bi}. Near threshold the the total photomeson cross section is dominated by the emission of 
single pions.  The most significant channel to consider involves the intermediate
production of the $\Delta$ resonance ~\cite{Stecker:1968uc}:

\begin{equation} 
p + \gamma \rightarrow \Delta \rightarrow N + \pi 
\end{equation}
This channel strongly dominates the photomeson production process near 
threshold. Since the UHECR flux falls steeply with energy, it follows that the bulk of the pions 
leading to the production of neutrinos will be produced close to the threshold.   

For a proton interacting with the CBR, a pion and a nucleon are produced. The outgoing nucleon has probability of 2/3 to be a proton and 1/3 probability to be a neutron from isospin considerations.  Should the resulting nucleon be a neutron, then the resulting pion is a $\pi ^{+}$.  Thus approximately twice the number of neutral pions are produced relative to charged pions from resonant pion production.  However direct pion production, which accounts for about 20\% of the total cross section, produces charged pions almost exclusively meaning that all told, approximately equal numbers of neutral and charged pions are produced around threshold.   Neutral pions decay into photons so one only need only consider the charged pions for neutrino production. Three neutrinos of roughly equal energy result from the decay chain of the $\pi ^{+} \rightarrow \mu^{+}\nu_{\mu} \rightarrow e^{+}\bar{\nu{_\mu}}\nu_{e}$.  

It is straightforward to determine the neutrinos produced and their energies from the ultrahigh energy cosmic rays.  We follow closely the calculation of the neutrino flux as described in reference \citep{Stecker:1978ah} and references therein.  The key is to determine the amount of energy that is carried away by the pion.   This follows directly from the inelasticity and the incident proton energy calculated using equation (\ref{pionproduction}).   For simplity, we assume here that all of the sources have the same primary injection spectrum and distribution.  One can then calculate the total neutrino flux by integrating over proton energy, photon energy, and redshift, assuming the standard $\Lambda$CDM cosmology.        

The effect of LIV on the photomeson neutrino production is manifested through the modification of the inelasticity of the interaction, since this determines the amount of energy that is carried away by the pion and therefore the resultant neutrino energy. The biggest impact of including LIV is to suppress the production of the higher energy photopions and therefore the resulting higher energy neutrinos. 

\begin{figure}
\begin{center}
\includegraphics[height=3.2in]{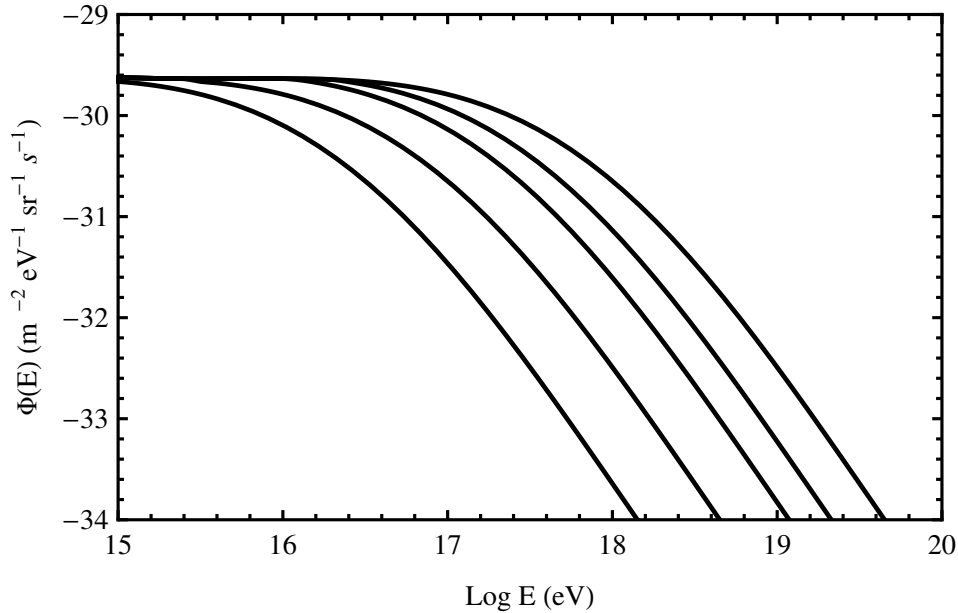}
\end{center}
\caption{Neutrino fluxes (of all species) corresponding to the UHECR models considered in Figure \ref{Auger}. From left to right, the curves give the predicted fluxes for $\delta_{\pi p} = 1 \times 10^{-22}, 6 \times 10^{-23}, 3 \times 10^{-23} , 1 \times 10^{-23}, 0$.~ \protect \cite{Scully:2010iv}}
\label{neutrinoflux}
\end{figure}

Figure \ref{neutrinoflux} shows the corresponding total neutrino flux (all species) for the same choices of $\delta_{\pi p}$ as the UHECR spectra presented in figure \ref{Auger}.  As expected, increasing $\delta_{\pi p}$ leads to a decreased flux of higher energy photomeson neutrinos as the interactions involving higher energy UHECRs are suppressed \cite{Stecker:2009hj}.  
 
\newpage

\section{Stable Pions from LIV}
\label{pi}

Almost all neutrinos are produced by pion decay. It has been suggested that if LIV effects can prevent the decay $\pi \rightarrow \mu + \nu$  of charged pions above a threshold energy, thereby eliminating higher energy neutrinos at the Glashow resonance energy and above~\cite{Anchordoqui:2014hua,Tomar:2015fha}. In order for the pion to be stable above a critical energy $E_{c}$, we require that its effective mass as given by equation (\ref{effectivemass}) is less than the effective mass of the muon that it would decay to, i.e., $\tilde{m}_{\pi} < \tilde{m}_{\mu}$. This situation requires the condition that $\delta_{\pi} < \delta_{\mu}$~\cite{Coleman:1998ti}. Then, neglecting the neutrino mass, this critical energy energy is given by
\begin{equation}
E_{c} = {{\sqrt{m_{\pi}^2 - m_{\mu}^2}}\over{\delta_{\mu} - \delta_{\pi}}}   
\end{equation}
\noindent noting that if, as before, we write $\delta_{\mu\pi} \equiv \delta_{\mu} - \delta_{\pi}$, then for small $\delta_{\mu\pi}$, it follows that $\sqrt{2\delta_{\mu\pi}} \simeq  \delta_{\mu\pi}$. In terms of SME formalism with Planck-mass suppressed terms, $\delta_{\mu\pi}$ is given by equation (\ref{d}) or equation (\ref{sub}). For example,
if we set $E_{c}$ = 6 PeV, in order to just avoid the Glashow resonance, we get the requirement,
$ \delta_{\mu\pi} \simeq 1.5 \times 10^{-8}$. However, it should be noted that a Glashow resonance event has been recently reported in Ref.~\refcite{IceCube:2021rpz}.

\section{Observational tests with new neutrino telescopes}
\label{tel}

Future neutrino detectors are being planned or constructed (see chapters 6 and 7): IceCube-Gen 2~\cite{IceCube-Gen2:2020qha}, the Askaryan effect detectors {\it ARA}~\cite{Guetta:2016bpa} and {\it ARIANNA}~\cite{Barwick:2016mxm}, and space-based telescopes such as {\it OWL}~\cite{Stecker:2004wt}, {\it EUSO}~\cite{Fenu:2017xct}, and a more advanced OWL-type instrument called {\it POEMMA}~\cite{POEMMA:2020ykm}. They will provide more sensitive tests of LIV.

\section*{Acknowledgments} 
I would like to acknowledge my collaborators: Stefano Liberati, David Mattingly and Sean Scully. I thank Francis Halzen for helpful comments on the IceCube neutrino telescope. I also thank Maria Gonzalez-Garcia and Michele Maltoni for helpful comments.

\newpage

\begin{appendix}[Appendix A: Dimensional Analysis with Mass Dimensions]
%\subsection*{Appendix: Dimensional Analysis with Mass Dimensions}
It is routine in particle theory to use "natural" units, setting $\hbar$ = $c = 1$.
Thus, in natural units, the dimensions of $\hbar$ and $c$ are then $[\hbar] = [c] = 0$. One defines the dimension of mass $[m] = +1$. It then follows that $[E] = [p] = +1$,
and $[t] = [x] = -1$. The derivative operator $\partial_{\mu}$ then has dimension
$[\partial_{\mu}]$ = +1.

The time evolution of a quantum system is described by a unitary transformation,
$e^{iS}$ where $S$ is the dimensionless action (remembering $[\hbar] = 0$), given
by the 4-integral of the Lagrangian density (or "Lagrangian"), $\cal{L}$,
\beq
\\\\\\\\\\\\S = \int{d^{4}}x\ \cal{L}\\\\\\\\\\\\\
\nonumber
\eeq
so that, since $[S] = 0$, $[\cal{L}]$ = +4. Scalar fields, $\phi$, 
scale inversely with $x$ so that [$\phi$] = +1. Fermion fields have $[\psi]$ = +3/2. 

Since $[\cal{L}]$ = +4, any operator consisting of fields adding up to $[\cal{O}]$ = +(4+n) must be balanced by a mass term $M^{-n}$, where $M$ is the scale of the EFT.
Thus, for the Fermi EFT with the 4-fermion operator $G_{F}[\psi][\psi][\psi][\psi]$, the Fermi coupling constant $G_F$ must have mass dimension $[G_F]$ = -2. In fact, its value is 1.166 $\times 10^{-5}$ GeV$^{-2}$ and is related to the weak mass scale, $M_W^{-2}$.
\end{appendix}

\begin{appendix}[Appendix B: Standard Model Extension Isotropic Diagonalizable Terms]

The standard model extension (SME) formalism restricted to diagonalizable and isotropic terms
yields a very simple class of models~\cite{Kostelecky:2011gq}
These models have simultaneously diagonalizable Lorentz-violating operators 
and they also are manifestly rotationally invariant in a preferred frame.
Thus, they can be related to the formalism used in Section \ref{form}.

For these models,
it is convenient to work in the diagonal basis
and in the preferred frame, however, see footnotes
in Section \ref{SME}.
In the diagonal basis and preferred frame,
the energy of a neutrino specific neutrino species, 
assuming the dominance of operators of even dimension $d$ with $\cal{CPT}$ even, 
can be written in the form
\begin{equation}
\Eri_{\nu} = {|{\bf p}|} + \frac{m_{\nu}^2}{2|{\bf p}|} 
+ \sum_{d} |{\bf p}|^{d-3}\,\ccfc{d}.
\label{even}
\end{equation}
In general, three coefficients for Lorentz violation appear for each $d$, 
one for each neutrino species. For relativistic neutrinos, $|{\bf p}| \gg m_{\nu}$
and the second term in equation (\ref{even}) can be neglected. See reference~\refcite{Kostelecky:2011gq}
for a general treatment including $\cal{CPT}$-odd terms and anisotropic terms.

\end{appendix}

\newpage

\bibliographystyle{ws-rv-van}
%\bibliography{bib,gc,frontiers} 
 
%\bibliographystyle{apsrmp4-1}
\bibliography{arXivLIV_ch}

%\end{chapter}
\end{document}